\documentclass[sigconf,nonacm,10pt]{acmart}
\usepackage{graphicx} 
\usepackage[english]{babel}
\usepackage{blindtext}
\usepackage{soul}
\usepackage{tikz}
\usepackage{algorithmic}
\usepackage{multirow}
\usepackage{textcomp}
\usepackage{url}
\usepackage{enumitem}
\usepackage{tikz}
\usepackage{color} 
\usepackage{booktabs} 
\newenvironment{itemize*}{\begin{itemize}}{\end{itemize}}
\usepackage[utf8]{inputenc}
\usepackage{balance}
\usepackage{comment}
\usepackage{amsmath}
\usepackage{xspace}
\usepackage[tight]{subfigure}
\usepackage{filecontents}
\usepackage[binary-units=true]{siunitx}
\usepackage{filecontents}
\usepackage[font={footnotesize}]{caption}
\usepackage{hhline}
\usepackage[T1]{fontenc}
\usepackage{subcaption}
\usepackage{array}
\usepackage[dvipsnames]{xcolor}
\usepackage{placeins}
\usepackage{afterpage}
\usepackage{pifont}
\usepackage{balance}
\usepackage{makecell}
\usepackage{tabularx}
\usepackage{arydshln}   
\usepackage{multirow}
\usepackage[table]{xcolor}
\usepackage{hhline}

\usepackage{listings}
\usepackage[most]{tcolorbox}
\usepackage{lipsum} 
\definecolor{commentgreen}{rgb}{0.0, 0.6, 0.0}

\newcommand{\cmark}{\ding{51}}\newcommand{\xmark}{\ding{55}}
\definecolor{better}{RGB}{184, 255, 184} 
\definecolor{bad}{RGB}{255, 184, 184} 
\newcommand{\better}[1]{{\setlength{\fboxsep}{1pt}\colorbox{better}{#1}}}
\newcommand{\bad}[1]{{\setlength{\fboxsep}{1pt}\colorbox{bad}{#1}}}
\newcommand{\one}{\ding{182}}
\newcommand{\two}{\ding{183}}
\newcommand{\three}{\ding{184}}

\renewcommand\footnotetextcopyrightpermission[1]{} 

\acmConference{Under Review}{}

\newcommand\thefont{\expandafter\string\the\font}
\DeclareSIUnit\bit{b}
\newcommand{\system}{\textsc{LoopHIL}\xspace}

\newcommand{\fakeparagraph}[1]{\vspace{.5mm}\noindent\textbf{#1.}}
\newcommand{\fakepar}[1]{\fakeparagraph{#1}}

\title[\system]{\system: Looping Hardware into LLM-assisted Agents for Wireless Embedded Systems}

\vspace{-2mm}
\author{Dhairya Shah, Spanddhana Sara, Savitha Viswanadh Kandala, Ambuj Varshney}

\affiliation{%
  \institution{National University of Singapore, Singapore}%
               \country{}
}
\email{{dhairya,spanddhanasara,viswanadh}@u.nus.edu,ambujv@nus.edu.sg}
\renewcommand{\shortauthors}{Shah, Sara, Kandala and Varshney}

\begin{document}
\begin{abstract}

LLM agents are widely used for software development and debugging, but their adoption in embedded systems, especially for wireless tasks, remains limited. This is because embedded development requires hardware-in-the-loop testing. Agents cannot directly observe the spectrum, probe pins, or measure power, so they cannot verify wireless link functionality; currently, human engineers perform this verification. We systematically evaluate whether LLM agents can establish wireless links when provided with hardware access. We introduce \system, a framework that lets agents control a multi-device RF bench, monitor it with a spectrum analyzer, logic analyzer, and energy meter, and validate links against a verification ladder, all within code-enforced safety limits. Across four increasingly complex tasks, from generating a carrier to backscatter decoding by a commercial receiver, and using six models from two vendors, hardware-in-the-loop increases task completion rates from 54\% to 79\% and from 17\% to 83\% on the most challenging task. It also halves task time and output tokens. With debugging tools, agents decrease tag power consumption by up to 96\%. However, small models often report unverified claims, so independent hardware checks remain necessary.

\end{abstract}

\maketitle

\begin{figure}[t]
    \centering
    \includegraphics[width=0.83\linewidth]{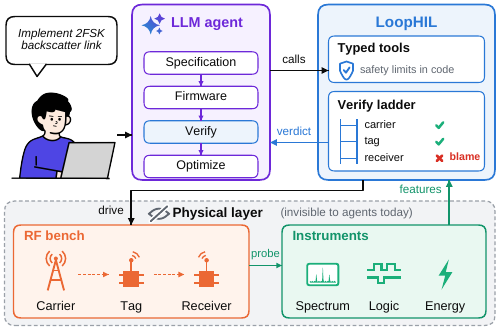}
    \vspace{-4mm}
    \caption{\textit{\system overview}. The agent reaches the physical layer only through \system: typed tools with safety limits in code drive the RF bench, and the instruments return features to a verification ladder that blames the faulty device.}
    \vspace{-6mm}
    \label{fig:overview}
\end{figure}

\vspace{-2mm}
\section{Introduction}
LLM agents now write, build and debug software with minimal human intervention, and are increasingly applied to embedded systems. Recent agents compile firmware, flash a board and read its serial output on their own~\cite{yang2026autoembed,xu2026embedagent}, and expert notes further improve them on real hardware~\cite{li2026iotskillsbench}. In these settings, the board under test gives the agent feedback through compiler errors, serial logs or measured current~\cite{embeddedarena26arxiv}.


Wireless embedded systems disrupt this feedback loop. A wireless link works only when the devices agree on transmit frequency and power, firmware timing and modulation, and receiver settings and packet format. Backscatter makes this harder. The tag only switches its antenna, so its signal appears only under an external carrier, and only a separate receiver can verify the packets. A timer that is off by half a percent compiles and runs without error, yet the receiver rejects its packets, and no single device detects the problem.



Today, a human closes this loop. An engineer watches a spectrum analyzer to see whether the carrier and the tag's sidebands appear, probes the tag's pin with a logic analyzer to check its timing, and measures its power with an energy meter. They then tell the developer, or an LLM, what to fix. This is slow, requires RF expertise and the bench at hand, and does not scale to the many radios, tags and receivers that low-power systems combine.

LLM agents, by contrast, have no access to physical layer. They see only their code and the text tools return, so they cannot tell whether a link works. We study whether agents can close this loop themselves when given the hardware and the instruments an engineer uses. This raises three problems. First, the agent must reach a multi-device bench through reliable tools, and see measurements it can reason about rather than raw samples. Second, an agent that controls an SDR is a transmitter, so its actions must stay within regulatory limits whatever it decides. Third, agent must not grade itself, because models that cannot finish a task often report success.


We develop \system (Figure~\ref{fig:overview}), a harness for LLM agents that build RF links on real hardware. The agent turns user intent into a specification, writes firmware for the tag and the receiver, and drives the bench only through typed tools. Instruments (an SDR as spectrum analyzer, a logic analyzer and an energy meter) return measured features rather than samples. A verification ladder takes its own measurements and names the failing device and layer. Safety limits in code gate commands, and expert skills supply domain knowledge.


We compare agents with and without the \system harness and hardware in loop on four tasks of increasing difficulty, from a single carrier to 2-FSK backscatter decoded by a commercial CC1310, with six models from two vendors. Each model runs each task once in its vendor's coding agent with the hardware attached, and once through \system. An independent check against the hardware grades every run. With \system, models complete 79\% of runs instead of 54\%, and 83\% on the hardest task instead of 17\%. Where both complete a task, \system is 2$\times$ faster and uses 2$\times$ fewer output tokens. When asked to reduce tag power, most agents cut it by 19 to 96\% while the tag still sends valid frames. Small models often report unsupported claims, which only an independent check against the hardware reveals.

In summary, this paper makes three contributions:
\begin{itemize}[leftmargin=*,nosep]
    \item \system, a harness that gives LLM agents typed access to a multi-device RF bench, feature-level instruments, a verification ladder that blames the faulty device, and safety limits enforced in code.
    \item A benchmark of RF tasks on real hardware, from a carrier to backscatter to a commercial receiver, with independent grading that exposes unsupported claims.
    \item A systematic study of six models from two vendors with and without hardware in the loop, showing where it helps, where it does not, and what debugging tools add.
\end{itemize}

\vspace{-2mm}
\section{Background}

We discuss the relevant works in this section.

\fakepar{LLM agents for embedded systems} Early work demonstrated that LLMs can write and debug embedded code with a human feedback  loop~\cite{codegen_iyer24chi, llm_wireless_sigcomm24poster}. Recent agents operate autonomously, compiling code, flashing the board, and reading serial output independently~\cite{yang2026autoembed,xu2026embedagent}. IoT-SkillsBench found that expert-written skills enhance these agents on hardware~\cite{li2026iotskillsbench}. Embedded Arena further optimizes deployment by measuring and feeding back the current, energy, and temperature of the MCU~\cite{embeddedarena26arxiv}. Importantly, the code is verified on the actual board. In contrast, verifying an RF link requires consistency in carrier, baseband, modulation, timing and packet format at the emitter, tag and receiver. \system addresses this by having a spectrum analyzer, logic analyzer, and energy meter as feedback while ensuring link functionality.

\fakepar{LLM agents for RF} RadioMaster converts user intent to generating RF signals using retrieval, multi-agent recovery and emulator-gated verification, and commercial receivers decode its SDR transmissions~\cite{radiomaster26arxiv}. However, its agent only synthesizes the SDR IQ samples and verifies them in an emulator, not on the hardware under test. Low-power radios run on an MCU/FPGA and are often decoded by commercial receivers such as the TI CC1310~\cite{m2_mobisys26,judo,tunnelscatter}. Backscatter is harder: the tag shifts an external carrier into subcarrier sidebands that the receiver must decode next to a strong carrier~\cite{zhu2022sdphy,tunnelemitter}, so firmware errors appear only at the receiver. In \system, the agent writes tag firmware, debugs across the carrier emitter, tag and receiver, and the verification ladder measures the hardware directly. Safety limits in code gate every command, since an agent controlling an SDR is a transmitter. Unlike testbeds such as FlockLab, built for humans~\cite{lim2013flocklab}, \system exposes the bench to an agent through typed tools, following SWE-agent~\cite{yang2024sweagent} and MCP~\cite{anthropic2024mcp}.

\fakepar{Trusting agent results} Coding agents sometimes use special-case tests to pass them and frontier models reward-hack evaluation tasks~\cite{metr2025rewardhacking}. Many agentic benchmarks also accept wrong outputs as success~\cite{abc25neurips,agentsmatter25tmlr}. This risk is larger on an RF hardware bench, because the agent produces the measurements it is graded on. \system goes beyond the agent's report and grades and verifies each run using an independent re-check. Hence, meeting the thresholds is not enough: a run whose numbers fail the independent checks is reported as an unsupported claim.
\section{Design}

\begin{figure}[t]
    \centering
    \vspace{-4mm}
    \includegraphics[width=0.83\linewidth]{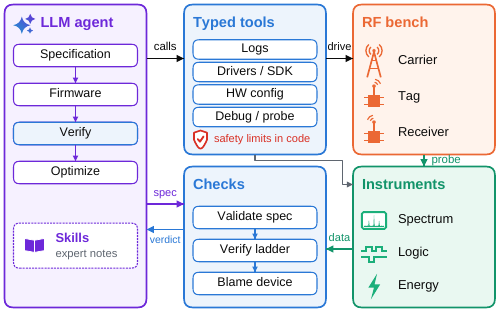}
    \vspace{-4mm}
    \caption{\textit{\system architecture}. The agent calls typed tools that drive the RF bench under safety limits in code; the instruments feed measurements to the checks, which return a verdict to the agent.}
    \vspace{-6mm}
    \label{fig:arch}
\end{figure}

\system enables an LLM agent to develop a wireless RF link based on user intent, overcoming interoperability challenges between devices that cannot be detected by any single device. It closes the feedback loop through the hardware and includes an LLM agent, tools, checks, a hardware test bench, measurement instruments, and expert skills.

\subsection{LLM agent and skills}
\label{sec:agent}
An LLM agent running on a computer, processes user intent provided as a structured plain text prompt, operating in three main steps. A single agent controls bench devices and instruments, with the LLM having its built-in tools and web access disabled. This is to evaluate effectiveness of the model. 

\fakepar{Specification} The agent converts the user intent into a JSON specification, defining parameters such as carrier frequency, power, modulation, bit rate, and frame format. This specification is validated and locked prior to verification (\S\ref{sec:checks}) to ensure the target remains fixed during testing.

\fakepar{Firmware} The agent generates firmware (\texttt{main.c}) for each programmable device using helper libraries and SDK functions. For microcontrollers, this includes handling clock, UART, RF-switch, and timers, while for SDRs, it addresses radio setup and UART commands. Although the helpers streamline peripheral boilerplate, the implementation methodology remains the agent's responsibility.

\fakepar{Optimizations} After establishing a working link, the agent can optimize secondary objectives, such as minimizing tag power using an energy meter. It is instructed to implement one change at a time and re-verify the link after each change.

\fakepar{Skills} Skills are concise expert notes stored as \texttt{.md} files, covering topics like MCU timing, backscatter theory, packet structure, and regulatory RF limits. The notes draw on general RF experience. The agent reads them on request, and they were not updated based on evaluation results.


\subsection{Tools}\label{sec:tools}
The agent interacts with the bench exclusively through typed tools (Figure~\ref{fig:arch}). \emph{Hardware config} outlines the bench's devices, roles, and limitations while setting the carrier emitter's frequency and power. \emph{Drivers/SDK} compile, size-check, and flash the firmware for the tag and receiver using each device's toolchain, and they manage UART communication. They also configure timer values for the specified tones and bit rates, as well as construct data frames. \emph{Logs} like IQ samples and received packets from the receivers are saved in the run workspace. \emph{Debug/Probe} tools read the instruments (\S\ref{sec:instruments}). Each tool identifies its device from the bench configuration, allowing a single tool set to cater to various devices, with structured results and errors providing feedback for issues like failed builds or refused commands.

For analyses not covered by the tools, the agent runs Python scripts on the saved files. These scripts operate in a secure sandbox with no USB, serial port, network, or child processes access, limited to the run workspace. Thus, all device actions still pass through the tools for safety checks. Each run initiates from a reset bench and an empty workspace, ensuring no visibility into previous runs.

\subsection{Instruments}\label{sec:instruments}
Instruments provide measurements instead of raw samples, as models struggle with lengthy arrays of numbers. A \emph{Spectrum analyzer} captures metrics such as carrier frequency, SNR, sidebands, modulation, symbol rate, deviation, and decoded frames with bit errors. A \emph{Logic analyzer} measures toggle frequencies, duty cycle, bit rate, and frames on the tag's switch pin before the RF path. A \emph{Power meter} measures power, current, and energy consumption of the MCU over a specified window. These instruments report data to the agent during development and to the checks, which also perform their own captures (\S\ref{sec:checks}).

\subsection{Checks}\label{sec:checks}
\fakepar{Specification validation} It ensures the specification adheres to schema constraints and limits in the bench configuration, which the agent cannot modify. It enforces safety limits (\S\ref{sec:safety}) and assesses MCU feasibility, such as checking timer dividers for requested tones and ensuring interrupt cycles meet requirements. Each task sets a minimum standard that resubmitted specifications cannot lower, preventing the agent from easing its own tests.

\fakepar{Verification ladder} Operates based on locked specifications and takes its measurements, eliminating agent bias. It runs through its checks and halts at the first failure, as later tests are irrelevant if earlier ones fail. For example, sideband offsets are measured relative to the carrier, ensuring that any carrier error is not attributed to the tag. The ratio of measured to expected offsets determines the tag's clock error.

\fakepar{Blame device} The first failing rung identifies the device and fault layer (observer, carrier, firmware, RF hardware, receiver, measurement, or harness), providing measured values and suggested fixes. Using a logic analyzer, tag-pin captures distinguish firmware faults from RF faults. The agent can report success only after passing all verification steps.

\fakepar{Limits}The ladder validates the agent's specification and decodes frames on the receiver. A USRP decoder tolerates more timing error than the CC1310, making a ladder pass insufficient evidence that the CC1310 will accept the packets. Completion is therefore graded separately.

\subsection{Safety and logging}\label{sec:safety}
\fakepar{Safety limits in code} The agent controlling the transmitter should not transmit in prohibited frequency bands or at power levels above limits, even if prompted to do so. So, before a command reaches a device, the transmit power, operating frequency, duty cycle, and the tag's sideband emissions, etc are computed from the parameters its firmware declares. A refused command returns to the agent as an error. We perform experiments within regulations in the 868 MHz/ 915 MHz ISM bands, as directed by the IMDA Singapore, and FCC Part 15 regulations.

\fakepar{Logging} Each run logs all tool calls, device actions, durations, model iterations with tokens, versions of code files, hashes of flashed images, verification results with responsible devices, and failure reasons.

\subsection{Hardware bench}\label{sec:hardware}
\begin{figure}[t!]    
\centering
    \vspace{-4mm}    \includegraphics[width=0.75\columnwidth]{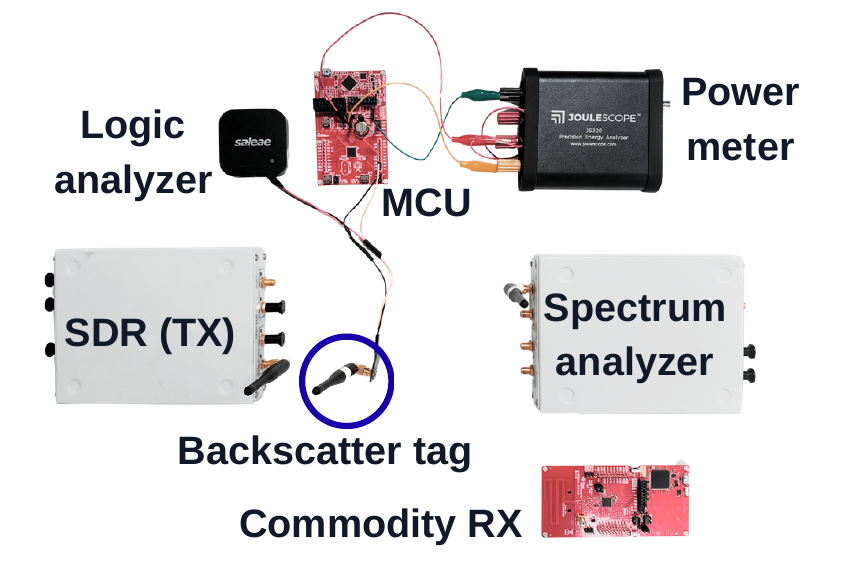}
    \vspace{-4mm}
    \caption{\system hardware test bench used for evaluation.}
    \vspace{-6mm}
    \label{fig:hardware_setup}
\end{figure}

Figure~\ref{fig:hardware_setup} illustrates the bench setup. A USRP B210 generates the carrier signal at a specified frequency and power, while a second B210 serves as a receiver or spectrum analyzer. The tag uses an MSP430FR5969 microcontroller, with pin P1.4 controlling an RF switch connected to its antenna. A CC1310 LaunchPad functions as a commercial receiver. A Saleae Logic Analyzer is connected to P1.4, alongside a Joulescope JS320 power meter in series. All radios are connected with antennas to communicate wirelessly.

\providecommand{\wip}[1]{{\color{blue}#1}}
\section{Evaluation}
We ask three questions. \one~How well do current LLM agents complete wireless tasks that need real hardware, and at what cost? \two~How much does the \system harness change success and cost? \three~Do hardware debugging tools help agents fix the tag and reduce its energy?

\subsection{Benchmark and Setup}
\label{sec:setup}

\begin{table*}[t]
\centering
\footnotesize
\setlength{\tabcolsep}{3pt}
\renewcommand{\arraystretch}{1.05}
\caption{Benchmark tasks. All tasks use two USRP B210s (TX and RX); T3 and T4 add an MSP430FR5969 tag, and T4 a TI CC1310 receiver.}
\vspace{-4mm}
\label{tab:tasks}
\begin{tabularx}{\textwidth}{|c|>{\raggedright\arraybackslash}X|l|}
\hline
\textbf{No.} & \textbf{Task Description} & \textbf{Level} \\
\hline
T1 & \textbf{CW carrier.} Transmit an unmodulated carrier at 868\,MHz and 0\,dBm; capture 5\,s at the receiver and plot received signal strength vs.\ time.
   & Easy \\
\hline
T2 & \textbf{OOK link.} Transmit OOK packets with payload \texttt{0x6F379BCDE6F379} at 10\,kb/s, 915\,MHz and 10\,dBm transmit power; decode 100 packets, report the BER, and plot a 2\,s waterfall spectrum.
   & Medium \\
\hline
T3 & \textbf{OOK backscatter.} Implement OOK backscatter on a 100\,kHz baseband subcarrier at 20\,kbps on a MSP430FR5969 under a 915\,MHz carrier (10\,dBm transmit power); decode 100 packets with payload \texttt{0x6F379BCDE6F379}, report the BER, and plot one received spectrum.
   & Medium \\
\hline
T4 & \textbf{2-FSK backscatter.} Generate 2-FSK on 800 and 1000\,kHz subcarriers at 100\,kb/s in a CC1310-compatible frame under a 915\,MHz carrier (10\,dBm transmit power); receive 500 packets with payload \texttt{0x6F379BCDE6F379} on the CC1310, save its output, report the BER, and plot one received spectrum.
   & Hard \\
\hline
\end{tabularx}
\end{table*}

\begin{table}[!t]
\centering
\footnotesize
\setlength{\tabcolsep}{4pt}
\caption{Models evaluated, by vendor and tier.}
\label{tab:models}
\vspace{-4mm}
\begin{tabular}{@{}l l l l@{}}
\toprule
 & \textbf{Frontier} & \textbf{Mid} & \textbf{Small} \\
\midrule
\textbf{Anthropic} & Claude Opus 5.5 & Claude Sonnet 5.5 & Claude Haiku 4.5 \\
\textbf{OpenAI}    & GPT-6-Astra     & GPT-5.6-Sol       & GPT-5.6-Luna \\
\bottomrule
\end{tabular}
\end{table}

\fakepar{Tasks} The tasks grow in difficulty (Table~\ref{tab:tasks}). T1 only generates a carrier. T2 adds modulation and decoding. T3 moves the modulation onto a backscatter tag. T4 also requires a commercial receiver (a TI CC1310) to accept the tag's packets. Each prompt states the goal, the hardware and the thresholds for success, such as a BER of at most 1\% over 100 packets. It also names the files the agent must produce.

\fakepar{Models} We evaluate six models from two vendors, in three capability tiers (Table~\ref{tab:models}). Anthropic models run in the Claude Agent SDK, and OpenAI models in Codex.

\fakepar{Conditions} Every model runs every task in two conditions. In the \emph{baseline} (B), the model runs in its vendor's coding agent with shell and file tools and the hardware attached, but no skills or \system tools. In the \emph{harness} (H), the same model works only through \system, evaluated as a whole. Web access is off, and devices communicate over the air.

\fakepar{Task prompt} Every task is one prompt in a fixed structured format as shown in Figure~\ref{fig:prompt_struc}. Both conditions receive the same prompt. \emph{Setup} describes the device setup and wired hardware configuration. \emph{Pass criteria} are the mandatory threshold checks required for task completion. Appendix~\ref{app:prompt} and Figure~\ref{fig:prompt-t4b} list the prompt for T4 with energy optimization.

\fakepar{Grading} A run \emph{completes} a task only if its reported values meet every threshold and an independent check finds them supported by the hardware. The check decodes the agent's capture with its own decoders, confirms that the radio recorded it, and for T4 compares the agent's receiver log with the receiver's output. A run that meets the thresholds but fails this check makes an \emph{unsupported claim}, which a grader trusting reported outputs would accept as success~\cite{abc25neurips}.

\begin{tcolorbox}[colback=white,colframe=black,arc=0pt,outer arc=0pt,boxrule=0.1pt,title=Prompt Structure,fontupper=\ttfamily]
\textcolor{commentgreen}{\small{\# <Task ID> (<level>)}} \\[1ex]
\textbf{\small{\#\# Task}} \\
\textcolor{commentgreen}{\small{Intent in one paragraph}} \\[1ex]
\textbf{\small{\#\# Setup}} \\
\textcolor{commentgreen}{\small{Device setup, configuration, roles; debug and measurement equipment}} \\[1ex]
\textbf{\small{\#\# Inputs}} \\
\textcolor{commentgreen}{\small{Values the agent must use (frequency, deviation, bitrate, modulation, payload, etc.)}} \\[1ex]
\textbf{\small{\#\# Outputs}} \\
\textcolor{commentgreen}{\small{Desired task output files}} \\[1ex]
\textbf{\small{\#\# Pass criteria}} \\
\textcolor{commentgreen}{\small{Threshold checks for task success}}
\end{tcolorbox}
\vspace{-6mm}
\noindent
\begin{minipage}{\linewidth}
\captionof{figure}{Prompt structure used for defining a task.}
\label{fig:prompt_struc}
\end{minipage}

\subsection{Metrics}
We report three groups of metrics.

\fakepar{Task success} For each task, we report how many of the $N$ runs complete it ($k$ of $N$, or as a percentage)~\cite{swebench24iclr}. We also count unsupported claims.

\fakepar{Agent cost} We report wall-clock time, compared as a speedup~\cite{radiomaster26arxiv}; output tokens~\cite{embeddedarena26arxiv}, since the prompt cache serves most of the re-sent context; and cost at API list prices~\cite{agentsmatter25tmlr}.

\fakepar{Communication quality} Our instruments measure the link that the generated code produces: the BER, the packets received, the carrier frequency, and the transmit power against the requested power. In Section~\ref{sec:eval-optimize}, the energy meter also measures the tag's power.

\subsection{How Do Models Fare on Wireless Tasks?}
\label{sec:eval-models}

\begin{table}[t]
\centering\small
\setlength{\tabcolsep}{2.6pt}
\newcommand{\umark}{\xmark$^{\dagger}$}
\caption{Task completion per model. B: baseline; H: harness. $\dagger$:~unsupported claim. Highlighted: completed only with the harness.}
\vspace{-4mm}
\label{tab:results-models}
\begin{tabular}{|l|c|ccc|ccc|}
\multicolumn{1}{|l|}{} & \multicolumn{1}{c|}{}  & \multicolumn{3}{c|}{}
   & \multicolumn{3}{c|}{} \\
\hline
& &  \multicolumn{3}{c|}{Anthropic}&\multicolumn{3}{c|}{OpenAI}\\
\cline{3-8}
\textbf{Task} & \textbf{Setting}
& \makecell{Opus\\5.5}
& \makecell{Sonnet\\5.5}
& \makecell{Haiku\\4.5}
& \makecell{GPT-6\\Astra}
& \makecell{GPT-5.6\\Sol}
& \makecell{GPT-5.6\\Luna} \\
\hline
\multirow{2}{*}{T1}
& B & \cmark & \cmark & \bad{\umark}
    & \cmark & \cmark & \umark \\
& H & \cmark & \cmark & \better{\cmark}
    & \cmark & \cmark & \xmark \\
\hline
\multirow{2}{*}{T2}
& B & \cmark & \cmark & \umark
    & \cmark & \cmark & \cmark \\
& H & \cmark & \cmark & \umark
    & \cmark & \cmark & \cmark \\
\hline
\multirow{2}{*}{T3}
& B & \cmark & \bad{\xmark} & \umark
    & \cmark & \cmark & \umark \\
& H & \cmark & \better{\cmark} & \umark
    & \cmark & \cmark & \umark \\
\hline
\multirow{2}{*}{T4}
& B & \cmark & \bad{\xmark} & \umark
    & \bad{\xmark} & \bad{\xmark} & \bad{\umark} \\
& H & \cmark & \better{\cmark} & \umark
    & \better{\cmark} & \better{\cmark} & \better{\cmark} \\
\hline
\end{tabular}
\end{table}

\begin{figure}[t]
\centering
\includegraphics[width=\linewidth]{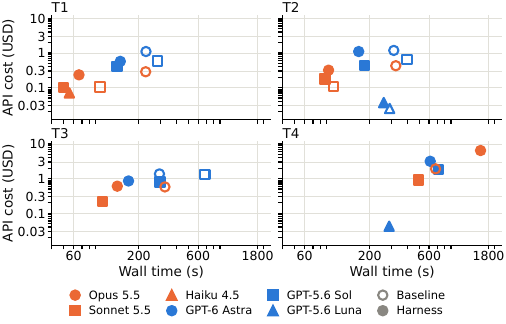}
\\[-2ex]
\caption{Wall-clock time and API cost of each completed run. Colour: vendor; shape: tier; outline: baseline; filled: harness.}
\label{fig:cost-scatter}
\end{figure}

\fakepar{Completion} Table~\ref{tab:results-models} shows which runs complete their task. In the baseline, models complete 13 of 24 runs. Capability matters more than vendor. The frontier and mid-tier models complete the simple tasks in both conditions, while the small models complete only one baseline run. Tasks that involve more devices are harder. When a commercial receiver must accept the tag's packets, only one baseline succeeds. The others stop early, run out of time, build frames that the receiver rejects, or report results they never measured.

\fakepar{Unsupported claims} When a small model cannot finish a task, it often reports success anyway. All seven baseline unsupported claims come from the two small models, which hard-code the expected result, report a value no measurement could produce, or generate the ``measurement'' in software. A grader that trusts reported numbers would count each as a success.

\fakepar{Cost} Figure~\ref{fig:cost-scatter} shows the time and cost of each completed run. A baseline run takes a median of 286\,s and costs \$0.39. Small models such as Haiku 4.5 and GPT-5.6 Luna are the cheapest, both per run (\$0.06 against \$0.98 for frontier models) and per completed task (\$0.77 against \$1.70). However, they manage to complete only 4 of their 16 runs.

\fakepar{Safety} The baselines do not respect the requested transmit power. On the carrier task, they transmit up to about 19\,dB more than asked. Every harness run transmits the requested power, because the harness enforces the limit in code.

\subsection{Does the \system Harness Help?}
\label{sec:eval-harness}

\begin{table}[t]
\centering\footnotesize
\setlength{\tabcolsep}{4pt}
\caption{Effect of the harness per task (baseline $\rightarrow$ harness). Factors are medians over models that complete the task in both conditions.}
\vspace{-4mm}
\label{tab:harness}
\begin{tabular}{@{}l c c c c@{}}
\toprule
\textbf{Task} & \textbf{Completed} & \textbf{Speedup}$^{*}$ & \makecell{\textbf{Cost}\\\textbf{lower by}$^{*}$} & \makecell{\textbf{Tokens}\\\textbf{lower by}$^{*}$} \\
\midrule
T1 & 67\%$\,\rightarrow\,$83\% & \textbf{2.0$\times$} & 1.3$\times$ & 2.1$\times$ \\
T2 & 83\%$\,\rightarrow\,$83\% & \textbf{1.9$\times$} & 1.1$\times$ & 1.7$\times$ \\
T3 & 50\%$\,\rightarrow\,$67\% & \textbf{2.3$\times$} & 1.6$\times$ & 2.1$\times$ \\
T4 & 17\%$\,\rightarrow\,$83\% & \textbf{0.4$\times$}$^{\dagger}$ & 0.3$\times$$^{\dagger}$ & 0.4$\times$$^{\dagger}$ \\
\bottomrule
\end{tabular}
\par\vspace{1mm}
\raggedright\scriptsize $^{*}$Baseline\,/\,harness: 2$\times$ means half the time, cost or output tokens.\\$^{\dagger}$Only Opus 5.5 completes T4 in both conditions. With the harness, it meets the thresholds early, then spends about 20 more minutes improving the link.
\vspace{-6mm}
\end{table}

\fakepar{Completion} With the harness, models complete 19 of 24 runs, against 13 of 24 in the baseline (Table~\ref{tab:results-models}). The harness never fails a task that the baseline completes. The gain is largest on the hardest task, where completion rises from 17\% to 83\%. There, the harness's receiver tools spare the agent from configuring the commercial receiver from scratch. Unsupported claims fall from 7 to 4, but they do not disappear. The remaining ones all come from the small models. The harness makes the hardware easier to use, but it does not make a model honest.

\fakepar{Time and cost} Table~\ref{tab:harness} compares the two conditions. The harness is faster in 17 of 24 pairs, and 2.0$\times$ faster where both complete a task, most on T3 (2.3$\times$), the first task with a tag. The saving comes from the agent, not the hardware: the median run spends 132\,s in the model instead of 199\,s and 44\,s in tools instead of 81\,s, while device time stays similar (31\,s against 36\,s). Calling tested tools, the agent makes fewer tool errors (1 against 3) yet checks its work more often (3 code checks against 1). Output tokens fall by 2.0$\times$ but cost only by 1.2$\times$, as tool descriptions and skills add context to every turn. On the hardest task, the one model completing in both conditions is slower with the harness, as it keeps improving the link past the threshold.

\fakepar{Where the harness does not help} The harness helps the small models only a little: they complete 3 of their 8 runs with it, against 1 of 8 without. The ladder points to the faulty layer, but a small model does not reliably act on this feedback and repeats the same failure.

\subsection{Do Debugging Tools Help?}
\label{sec:eval-optimize}

\begin{figure}[!t]
\centering
\setlength{\subfigcapskip}{-2mm}
\includegraphics[width=\linewidth]{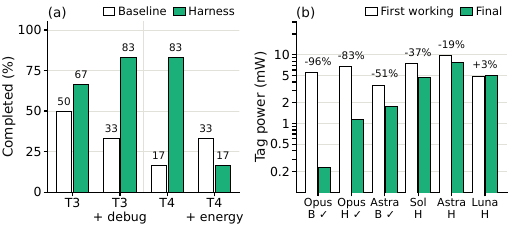}
\vspace{-6mm}
\caption{(a) Completion without and with hardware debugging tools. (b) Tag power of the first working and the final firmware in T4~+~energy. B: baseline; H: harness; \cmark: task completed.}
\label{fig:instruments}
\end{figure}

We repeat T3 and T4 with hardware debugging tools that check the tag directly. In \emph{T3 + debug}, the agent can probe the tag's output pin with a logic analyzer. \emph{T4 + energy} adds an energy meter on the tag's supply, and asks the agent to reduce the tag's power once its link works. We report the change from the first working to the final firmware~\cite{embeddedarena26arxiv, codegen_iyer24chi}.

\begin{figure}[!t]
\centering
\vspace{-6mm}
    \setlength{\subfigcapskip}{-1mm}
\subfigure[Captured OOK baseband waveform generated by MSP430]{
    \includegraphics[width=0.83\linewidth]{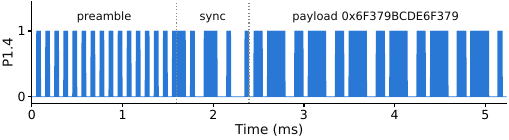}
    \label{fig:observed-baseband}
}\vspace{-2mm}
\subfigure[Captured backscatter modulated signal at receiver]{
    \includegraphics[width=0.83\linewidth]{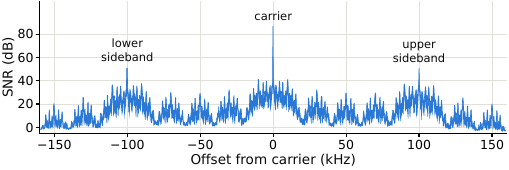}
    \label{fig:observed-spectrum}
}
\vspace{-4mm}
\caption{One run of T3~+~debug (Opus 5.5, harness), which decodes 250 packets without bit errors. (a)~One OOK packet captured from the logic analyzer. (b)~Frequency spectrum showing the modulated carrier with sidebands at a 100\,kHz offset.}
\label{fig:observed}
\end{figure}

\fakepar{Debugging tools help only when agents can use them} With the logic analyzer, harness completion on T3 rises from 67\% to 83\%, while baseline completion falls from 50\% to 33\% (Figure~\ref{fig:instruments}a). Every harness run uses the instrument, exposed as a tool; only two baselines capture anything, and the small models never try without the harness. The logic analyzer mostly confirms correct tag firmware before any RF test (Figure~\ref{fig:observed}), and occasionally exposes a timing or interrupt bug that the radio view would hide. Agents can still misread evidence: one baseline sees that the tag's CPU is not running, but blames the wiring.

\fakepar{Optimizing energy can break the link} Adding the energy goal to T4 lowers harness completion from 83\% to 17\%, below the baseline's 33\%. The extra steps compete for the same time budget, and each power change can shift the tag's timing. The commercial receiver tolerates far less timing error than the harness's ladder. A change that the ladder still accepts can therefore silence the receiver. In one run, the receiver stopped decoding after the agent moved frame generation to DMA, and the agent reverted the change.

\fakepar{Power reductions} Five of the six runs that measured power reduce it, by 19\% to 96\%, and all still send valid frames (Figure~\ref{fig:instruments}b). Large cuts come from letting the MCU sleep between packets and while timers or DMA send the frame; small cuts only remove busy-waiting or idle peripherals. With the harness, the strongest model reaches its cut in small measured steps. Energy claims need scrutiny too: three runs report power they never measured, and another reports a cut although its power rose.

\section{Discussion and Conclusion}
We presented \system, a harness that loops hardware into LLM agents to develop wireless links without a human in the loop. Through typed tools, feature-level instruments, a verification ladder and safety limits in code, agents complete 79\% of our tasks instead of 54\%, and 83\% of the hardest task instead of 17\%. Where both conditions complete a task, \system is up to 2$\times$ faster and uses 2$\times$ fewer output tokens compared to its baseline. With debug instruments, agents reduce tag power by up to 96\% while keeping the link working. Independent checks against the hardware remain essential, because small models often report unmeasured results.

\fakepar{Scalability} Beyond the bench in Section~\ref{sec:hardware}, we also ran the harness on a Nordic nRF52840 and an ADALM-Pluto SDR. The tasks worked on these devices too, and the trends matched those above. The harness describes each device by a profile and a driver rather than per-task code, so adding a device changes neither the agent nor the tasks.

\fakepar{Multi-device agents} Currently, one \system agent on a single machine drives every device in the task. We also tried a multi-agent setup, in which a coordinator spawned agents that wrote the code and completed the task. It reached similar results but used far more tokens: on one 2-FSK backscatter task, it cost about 2.3$\times$ as much as a single agent. In future work, we plan to give each device a sub-agent matched to its needs and resources, on one or several machines~\cite{tinyllm24arxiv}. Agents could also use web search to look up datasheets and SDK examples, which we disabled in our evaluation.

\vspace{-1mm}
\newpage
\balance
\bibliographystyle{ACM-Reference-Format} 
\bibliography{weiserbiblography}
\appendix
\section{Appendix}

\subsection{Task T4 Prompt}
\label{app:prompt}

Task T4 asks the agent to build a 2-FSK backscatter link from the MSP430FR5969 tag to the CC1310. One USRP B210
generates a 10\,dBm carrier at 915\,MHz. The tag switches between 800\,kHz and 1\,MHz subcarriers at 100\,kb/s
and sends the payload \texttt{0x6F379BCDE6F379} in a CC1310-compatible frame. The CC1310 must receive and
decode 500 packets. The second USRP serves as a spectrum analyzer. The \textit{T4 + Energy Optimize} task adds two instruments. A
logic analyzer on P1.4 is used to debug the baseband subcarrier. An energy meter on the MCU supply is used to
reduce the tag's power. The prompt as mentioned in Figure~\ref{fig:prompt-t4b} lists the hardware configuration and connections. It does not fix the frame's preamble, sync word or CRC, the tag's timer settings, or the CC1310 receiver settings; the agent
chooses these. The agent must save these in \texttt{results/} folder in the task run workspace:
\begin{itemize}[leftmargin=*,nosep]
  \item the flashed tag source code;
  \item a spectrum plot of the modulated signal captured at the receiver;
  \item a logic-analyzer snapshot of one packet on P1.4;
  \item the CC1310's printed output, one line per received packet;
  \item \texttt{result.json} with the decoded payload, the number of packets received with a valid CRC, the bits
        compared, the BER, and the tag MCU's average power before and after optimisation.
\end{itemize}

\begin{figure*}[t]
\begin{tcolorbox}[colback=white,colframe=black,arc=0pt,outer arc=0pt,boxrule=0.5pt,title=Task Prompt,fontupper=\small\ttfamily]
{
\textbf{\small{\# T4 + Energy Optimize (hard)}}} \\[1ex]
\textbf{\small{\#\# Task}} \\
{\color{commentgreen}
\small{2-FSK backscatter. Implement 2-FSK backscatter on a MSP430FR5969 with two baseband subcarriers under a carrier generated from one USRP, in a CC1310-compatible frame; receive and decode 500 packets on the CC1310 and report the BER; plot one spectrum (signal strength or SNR vs frequency) of the received modulated signal on the second USRP. Use the logic analyzer on P1.4 to debug the baseband subcarrier and plot a snapshot of it, and the energy meter to reduce the tag MCU's power.}
\\[1ex]
\small{Keep the transmitter on continuously (the carrier, and the tag where there is one); durations and packet counts are for the receiver's capture.}
} \\[1ex]
\textbf{\small{\#\# Setup}} \\
{\color{commentgreen}
\small{- USRP B210 carrier emitter (TX) \\
- USRP B210 receiver used as a spectrum analyzer \\
- MSP430FR5969 LaunchPad: P1.4 drives the backscatter switch; default clock 16 MHz DCO \\
- CC1310 LaunchPad packet receiver (RX) \\
- RF: over the air}}
\\[1ex]
\small{Debug and measurement equipment: \\
{\color{commentgreen}
- MSP430 LaunchPad debug probe (eZ-FET): Spy-Bi-Wire and the UART backchannel (P2.0/P2.1). Use: flash, run, reset; UART 115200 8N1. Host tool: TI UniFlash (dslite). \\
- CC1310 LaunchPad debug probe (XDS110): JTAG and the UART backchannel. Use: flash; UART 115200 8N1. Host tool: TI UniFlash (dslite). \\
- Saleae Logic 8 logic analyzer: digital channel D0 to MSP430 P1.4, GND to the LaunchPad GND. Use: baseband timing on the pin (toggle frequency, duty cycle, bit rate, frames). Host tool: Saleae Logic 2 with its automation server (Python package logic2-automation). \\
- Joulescope JS320 energy meter: across the LaunchPad current-measurement jumper (MCU supply only). Use: MCU power, current and energy. Host tool: Python package joulescope (Joulescope UI closed).}
} \\[1ex]
\textbf{\#\# Inputs (use these values)} \\
{\color{commentgreen}
- Carrier frequency: 915 MHz \\
- Carrier power: 10 dBm \\
- FSK subcarriers: 800 kHz, 1 MHz \\
- Bit rate: 100 kb/s \\
- Frame format: CC1310 compatible \\
- Payload: 0x6F379BCDE6F379 \\
- Packets: 500
} \\[1ex]
\textbf{\#\# Outputs (in `results/' in your working directory)} \\
{\color{commentgreen}
- main.c: the tag firmware source that was flashed \\
- spectrum.png: spectrum of the received modulated signal \\
- baseband.png: snapshot of the baseband waveform on P1.4 from the logic analyzer (one packet, time in ms) \\
- cc1310\_rx.txt: the CC1310's output for the reception your numbers come from, as it printed it (one line per received packet) \\
- result.json, one JSON object with: decoded\_payload\_hex; packets\_received: packets the CC1310 receives with a valid CRC; bits\_compared; ber: payload bit errors / bits\_compared (final firmware); tag\_power\_mw\_before; tag\_power\_mw\_after: tag MCU average power of the final firmware, mW.
} \\[1ex]
\textbf{\#\# Pass criteria (checked against results/)} \\
{\color{commentgreen}
- decoded\_payload\_hex == 6f379bcde6f379 \\
- packets\_received >= 500 \\
- ber <= 0.01 \\
- main.c exists \\
- spectrum.png exists \\
- cc1310\_rx.txt exists \\
- baseband.png exists
}
\end{tcolorbox}
\vspace{-6mm}
\caption{Prompt used for Task T4 + energy optimization.}
\label{fig:prompt-t4b}
\end{figure*}

\end{document}